\documentclass[epj]{svjour}
\usepackage{latexsym}
\usepackage{graphicx}
\usepackage[usenames,dvipsnames]{xcolor} 
\usepackage{amsmath,amssymb}

\begin{document}
\title{Can the induced increase in the angular velocity prevent the overspinning of BTZ black holes?}
\author{Koray D\"{u}zta\c{s}
}                     
%

\institute{Faculty of Engineering and Natural Sciences, \.{I}stanbul Okan University, 34959 Tuzla \.{I}stanbul, Turkey}

\date{Received: date / Revised version: date}
%
\abstract{Previously we showed that nearly extremal Ba\~{n}ados-Teitelboim-Zanelli (BTZ) black holes can be overspun by test bodies and fields, following the work of Rocha and Cardoso for the extremal case. The naked singularities in AdS space-times correspond to states rotating faster than light in the Ads/CFT correspondence. Therefore, overspinning turns out to be a drastic problem in a (2+1) dimensional AdS space-time, where one cannot invoke backreaction effects. Here, we consider the induced increase in the angular velocity of the event horizon which modifies the condition to allow the absorption of the perturbations satisfying the null energy condition. We show that its magnitude is sufficiently large to prevent the absorption of the challenging modes both for test bodies and scalar fields. We bring a solution to the notorious overspinning problem  which does not involve any reference to self-energy or gravitational radiation. 
%
\PACS{
      {04.20.Dw}{Singularities and cosmic censorship}   
     } 
} 
\maketitle
\section{Introduction}
The (2+1) dimensional black hole solution in the asymptotically anti-de Sitter background derived by Ba\~{n}ados, Teitelboim, and Zanelli (BTZ) have considerably intrigued researchers over the last three decades \cite{btzmain}. The BTZ black hole is a solution of vacuum Einstein equations with a negative cosmological constant $\Lambda=-1/\ell^2$. The metric is uniquely described by the mass and  angular momentum  parameters in accord with the no-hair theorem. 
\begin{equation}
ds^2=-N^2dt^2+N^{-2}dr^2+r^2(N^{\phi}+d\phi )^2
\label{btzmetric}
\end{equation}
Here $N$ and $N^{\phi}$ are the usual lapse and shift functions of the radial coordinate $r$:
\begin{equation}
N^2=-M + \frac{r^2}{\ell^2} + \frac{J^2}{4r^2}, \quad N^{\phi}=-\frac{J}{2r^2}
\end{equation}
where $M$ and $J$ denote the mass and the angular momentum parameters, as usual. The metric has two coordinate singularities which determine the locations of the inner and outer event horizons.
\begin{equation}
\frac{2r_{\pm}^2}{\ell^2}=M \left( 1 \pm \sqrt{1-\frac{J^2}{M^2 \ell^2}} \right)
\end{equation}
The existence of the event horizon relies on the condition that  the angular momentum of the black hole does not exceed the upper limit:
\begin{equation}
J \leq M \vert \ell \vert
\label{condibtz}
\end{equation}
The condition (\ref{condibtz}) resembles the condition $J\leq M^2$ for Kerr black holes to ensure the event horizon exists and the curvature singularity at $r=0$ is causally disconnected from distant observers at asymptotically flat infinity. This leads us to the notorious problem of cosmic censorship in classical general relativity. After proving that singularities inevitably ensue in gravitational collapse \cite{pensing}, Penrose conjectured these singularities to be hidden behind the event horizons which would not allow any effect propagating out of the singularities to reach distant observers \cite{ccc}. This way one can circumvent the existence of the singularity and the smooth structure of the space-time is preserved  at least outside the black hole region. As a concrete proof of cosmic censorship appeared to be elusive, Wald constructed an alternative process to check whether event horizons can be destroyed and hidden singularities can be exposed. In this type of thought experiments one starts with a black hole and attempts to increase its angular momentum or charge beyond the upper limit, by perturbing it with test particles or fields \cite{wald74}. After various attempts involving different scenarios it turns out that perturbations satisfying the null energy condition cannot destroy the event horizons \cite{hu,js,backhu,backjs,f1,gao,siahaan,magne,yuwen,higher,v1,he,wang,jamil,shay3,shay4,zeng,semiz,q1,q2,q3,q4,q5,q6,q7,overspin,emccc,natario,duztas2,mode,taubnut,kerrsen,kerrmog,hong,yang,bai,tjphys,khoda,ong,corelli1,molla,zhao}. This is due to the fact that there exists a lower limit for the energy of the perturbation to allow its absorption by the black hole, provided that the null energy condition is satisfied. The relative contributions of the test particle or field to the mass, angular momentum, charge parameters  satisfy: 
\begin{equation}
\delta M \geq \Omega \delta J + \Phi \delta Q
\label{needham}
\end{equation}
The first derivation of the relation (\ref{needham}) known to this author is dated back to Needham in 1980 \cite{needham}. Therefore we will refer to (\ref{needham}) as the Needham's condition. This condition prevents the absorption of modes with relatively high angular momentum or charge, which could potentially destroy  the event horizon.
For test fields with an energy-angular momentum ratio $(\delta M)/(\delta J)=m/\omega$, Needham's condition (\ref{needham}) reduces to $\omega \geq m\Omega$, which is the well-known super-radiance condition. The test fields with lower frequencies are reflected back to infinity with a larger amplitude. To be more precise the absorption probability for these fields is negative \cite{page}. However,  the absorption probability is always positive for fermionic fields the energy momentum tensor of which do not satisfy the null energy condition. A corresponding lower bound for the energy does not exist. The absorption of the low energy modes is allowed which could lead to an excess increase in the angular momentum or charge parameters resulting in the destruction of the event horizon  \cite{duztas,toth,generic,spinhalf,threehalves}. In \cite{threehalves} we have made an elucidative comparison of the cases that do and do not satisfy the null energy condition and pointed out that they should not be confused.

The main motivation to conjecture the singularities to be hidden behind the event horizons was to disable their causal contact with distant observers at asymptotically flat infinity. Though the notion of distant observers may not be well defined in the asymptotically de-Sitter and anti-de Sitter space-times, whether the singularities remain covered by event horizons is still an interesting problem. Therefore the analysis to test the possibility of destroying event horizons have also been extended to these cases \cite{rc,btz,gwak3,chen,ongyao,mtz,he2,dilat,yin,btz1,yang1,shay5,gwak4,shay6,corelli,sia2}. An alternative motivation for the anti-de Sitter case could be attributed to the AdS/CFT correspondence. The naked singularities in AdS space-time correspond to states at the conformal boundary rotating faster than the speed of light. Therefore one would like to prove that their existence is prohibited. With this motivation, Rocha and Cardoso attempted to prove that a BTZ black hole cannot be overspun past the extremal limit \cite{rc}. They started with an extremal BTZ black hole and showed that it cannot be overspun by a test body. Despite the fact that their analysis was restricted to the case of extremal black holes, they reached a general conclusion to state that BTZ black holes cannot be overspun. Furthermore, they claimed that one should be distrustful of any process that leads to overspinning in AdS, since this would refer to the existence of states rotating faster than light in the AdS/CFT correspondence. However in \cite{btz}, we showed that the line of research developed by Rocha and Cardoso themselves for test bodies leads to overspinning of BTZ black holes if one starts with a nearly extremal black hole, instead of an extremal one. We also analysed the case of test fields and derived the same results; i.e. though extremal black holes cannot, nearly extremal black holes can be overspun by test fields. 

The overspinning of nearly extremal black holes occurs via a discrete jump. The black hole cannot be continuously driven to extremality and beyond as indicated by the fact that extremal black holes cannot be overspun. In general, such an overspinning would not be generic and it would be fixed by backreaction effects. However, the case of (2+1) dimensions is problematic regarding the employment of backreaction effects. There are no gravitational-wave degrees of freedom so gravitational radiation cannot be evoked to prevent overspinning. The possibility to use self energy also remain controversial. This led Rocha and Cardoso to conclude that if overspinning were possible while one cannot invoke self energy or gravitational radiation, this would be the end of the story. In this work we bring a solution to this notorious problem. We consider the induced increase in the angular velocity of the event horizon described by Will  in his seminal work \cite{will}. We check whether this would be sufficient to prevent the absorption of the challenging modes for both test particles and fields and solve the overspinning problem without employing back-reaction effects.

We first re-state the overspinning problem derived by Rocha-Cardoso and D\"{u}zta\c{s} in \cite{rc,btz}, respectively. Then we carry out a more detailed analysis to determine the range of the challenging modes for test particles and fields. We calculate the induced increase in the angular momentum for the challenging modes, to check whether it is sufficiently large to prevent overspinning.

\section{Overspinning problem for BTZ black holes}
To test the validity of cosmic censorship for BTZ black holes, Rocha and Cardoso considered a black hole with initial parameters $M_0$ and $J_0$. They defined \cite{rc}:
\begin{equation}
j_0 \equiv \frac{J_0}{M_0 \ell} \leq 1
\end{equation}
They envisaged a test particle incident on the black hole with mass $m_0$, angular momentum $\delta J=m_0 L$ and energy $\delta M=m_0 E$. They derived the geodesic equations and they found the lower limit for the energy by imposing the geodesic to be future directed.
\begin{equation}
\frac{L}{E}\leq \frac{2r_+^2}{J}
\label{condirc}
\end{equation}
The condition (\ref{condirc}) should be satisfied to ensure the test particle is absorbed by the black hole. Notice that 
(\ref{condirc}) is identically equal to the Needham's condition (\ref{needham}) for the neutral case:
\begin{equation}
\delta M \geq \Omega \delta J
\label{needhamneut}
\end{equation}
where $\Omega=J/(2r_+^2)$ is the angular velocity of the event horizon for BTZ black holes. Rocha and Cardoso proceed to Taylor expand the final value of $j$ to check whether it can be larger than 1.
\begin{equation}
j=j_0 -\frac{\delta M}{M_0} \left( j_0 - \frac{L}{E \ell} \right)
\label{jfinrc}
\end{equation}
For an extremal black hole $j_0=1$ by definition. The condition (\ref{condirc}) implies that $(L/E)\leq \ell$. Therefore $j \geq 1$ for extremal black holes. Extremal BTZ black holes cannot be overspun.

As we mentioned above Rocha and Cardoso only evaluated the case of extremal black holes and fallaciously reached a general conclusion to state that BTZ black holes cannot be overspun. In \cite{btz} we followed the line of research of Rocha and Cardoso adapting the same notation. The only difference is that we started with a nearly extremal black hole instead.
\begin{equation}
j_0=\frac{J_0}{M_0 \ell}=1 -2\epsilon^2
\label{parambtz}
\end{equation}
where $\epsilon \ll 1$ parametrizes the closeness to extremality. This implies
\begin{equation}
\frac{2r_+^2}{\ell^2}=M_0 (1+2\epsilon)
\end{equation}
For a nearly extremal BTZ black hole the condition (\ref{condirc}) takes the form:
\begin{equation}
\frac{L}{E} \leq \ell (1+2\epsilon +2\epsilon^2)
\end{equation}
Choosing the maximum value for $L/(E \ell)$, we can evaluate (\ref{jfinrc})
\begin{equation}
j=1-2\epsilon^2 +(2\epsilon +4 \epsilon^2)\frac{\delta M}{M_0}
\end{equation}
We see that choosing $\delta M=M_0 \epsilon$ leads to $j>1$, i.e. the BTZ black hole is overspun beyond extremality \cite{btz}. Adapting the line of research developed by Rocha and Cardoso to nearly extremal black holes results in overspining. This appears to be a drastic result in the (2+1) dimensional case where one cannot employ back-reaction effects. 
\section{The induced increase in the angular velocity}
In this section we re-formulate the overspinning problem to enable the incorporation of the induced increase in the angular velocity of the event horizon into the analysis. We start with a nearly extremal black hole satisfying:
\begin{equation}
\frac{J^2}{M^2 \ell^2}=1-\epsilon^2
\label{param1}
\end{equation}
Notice that we made a slight change of notation, here. We do not use the subscript ``zero'' for the initial parameters. We also omitted the factor ``2'' appearing in (\ref{parambtz}).  
The parametrization (\ref{param1}) implies
\begin{equation}
j_0=\frac{J}{M\ell}=1-\frac{\epsilon^2}{2},  \quad \frac{2r_+^2}{\ell^2}=M(1+\epsilon)
\label{param2}
\end{equation}
We can calculate the angular velocity of the event horizon for a nearly extremal BTZ black hole:
\begin{equation}
\Omega =\frac{J}{2r_+^2}=\frac{1}{\ell}\left( 1-\epsilon+\frac{\epsilon^2}{2} \right)
\label{omegah}
\end{equation}
The upper bound for the angular momentum of the perturbation is given by the Needham's condition (\ref{needhamneut}):
\begin{equation}
\delta J \leq \frac{\delta M}{\Omega}= \delta M \ell (1+\epsilon +\epsilon^2/2 )
\end{equation}
The perturbations with a relatively higher angular momentum and lower energy are not absorbed by the black hole. The perturbations at the upper bound that saturate the Needham's condition, are referred to as optimal perturbations. Now, we check whether  optimal perturbations lead to overspinning in case one cannot evoke backreaction effects.
\begin{eqnarray}
j&=& j_0-\frac{\delta M}{M} \left( j_0 - \frac{\delta J}{ \delta M \ell} \right) \nonumber \\
&=& 1- \frac{\epsilon^2}{2}+\frac{\delta M}{M} ( \epsilon + \epsilon^2 )
\label{j1}
\end{eqnarray} 
with $\delta M =M\epsilon$, one observes that $j>1$, i.e. the BTZ black hole is overspun. We can also derive the lower bound for $\delta J$ for overspinning to occur from (\ref{j1}), which corresponds to the case $j=1$.
\begin{equation}
\delta J < \delta M \ell \left(1+ \frac{\epsilon}{2} -\frac{\epsilon^2}{2} \right)
\end{equation}
This determines the range of $\delta J$ for overspinning  to occur. To second order:
\begin{equation}
\delta M \ell \left(1+ \frac{\epsilon}{2} -\frac{\epsilon^2}{2} \right) < \delta J \leq \delta M\ell \left(1+\epsilon +\frac{\epsilon^2}{2} \right)
\label{range}
\end{equation}
The upper bound for $\delta J$ saturates the Needham's condition. We derived it by demanding that the test body is absorbed by the black hole. The lower bound is derived by demanding that $j>1$ in the final case, which indicates that overspinning has  occurred. Note that overspinning occurs for modes with $\delta J$ larger than the lower limit, and less than or equal to the upper limit in (\ref{range}).

Now we consider the induced increase in the angular velocity of the event horizon described by Will in his seminal paper \cite{will}. The induced increase in the angular velocity occurs due to the interaction of the test particle/field with the black hole before the absorption of the particle/field takes place. To avoid any confusion, the increase in the angular velocity of the event horizon does not imply that the angular momentum parameter of the space-time increases. Otherwise, overspinning would have occurred before the absorption of the test particle or field. The magnitude of the induced increase is derived by adding the angular momentum $\delta J$ of the perturbation to a black hole in the limit  $J \to 0$, and calculating the corresponding angular velocity. For a Kerr black hole this gives
\begin{equation}
\Delta \Omega =\frac{\delta J}{4M^3}
\end{equation}
with a corresponding self-energy for the test particle/field:
\begin{equation}
E_{\rm{self}}=\frac{(\delta J)^2}{4M^3}
\label{selfkerr}
\end{equation}
The induced increase in the angular velocity and the self-energy both act to prevent the overspinning of the black hole. The induced increase in the angular velocity prevents the absorption of the challenging modes by modifying the Needham's condition (or the absorption probabilities in a more subtle approach). The self energy contributes to the mass parameter of the space-time precluding the angular momentum parameter to exceed it in the final case. Recently, we calculated the Sorce-Wald condition \cite{sw} for the second order perturbations of a Kerr black hole and derived that \cite{spin2}
\begin{equation}
\delta^2 M - \Omega \delta ^2 J \geq \frac{(\delta J)^2}{4M^3}
\label{swcondi}
\end{equation}
Incorporating the self-energy given in (\ref{selfkerr}) with $(\delta ^2 J)=0$, is effectively equivalent to imposing the Sorce-Wald condition (\ref{swcondi}). The fact that the two approaches give the same result lends credence to the validity of the derivations. In \cite{spin2,absorp}, we argued that the conditions derived by Sorce and Wald in \cite{sw} are correct. However the function $f(\lambda)$ involves order of magnitude problems as one is forced to multiply the contributions of the second order perturbations by the square of the extra parameter $\lambda$. The order of magnitude problems in $f(\lambda)$ emerge as a brute algebraic fact. One can simply avoid these problems by abandoning $f(\lambda)$ and following the line of research developed by Semiz and D\"{u}zta\c{s}. This way, one does not have to multiply the contribution of the second order perturbations by $\lambda^2$ which would render it fourth order.

The employment of self-energy is could be a controversial subject in (2+1) dimensions. Therefore, we will restrict ourselves to the induced increase in the angular velocity of the event horizon. To calculate the induced increase in the angular velocity, we consider a BTZ black hole in the limit $J \to 0$ and consider the angular velocity it would acquire by perturbing it with a particle with angular momentum $\delta J$. This gives:
\begin{equation}
\Delta \Omega =\frac{\delta J}{2M \ell^2}
\label{indangvel}
\end{equation}
This value should be added to the angular velocity of the event horizon, which is modified as:
\begin{equation}
\Omega_{\rm{mod}}=\Omega +\Delta \Omega
\end{equation}
which  leads to the modification of the Needham's condition:
\begin{equation}
\delta M \geq \Omega_{\rm{mod}} \delta J
\label{needhammod} 
\end{equation}
When one takes the induced increase in the angular velocity of the horizon into account, the perturbations have to satisfy the modified form of the Needham's condition (\ref{needhammod}). This prevents the absorption of the low energy modes in favour of the cosmic censorship conjecture. The upper bound in (\ref{range}) saturates the Needham's condition. A slight increase in $\Omega$ would prevent the absorption of the modes at the upper bound. Therefore we should focus on the lower bound in (\ref{range}). For these modes the induced increase in the angular velocity is given by:
\begin{equation}
\Delta \Omega =\frac{\delta M \ell \left(1+ \frac{\epsilon}{2} -\frac{\epsilon^2}{2} \right) }{2M \ell^2}=\frac{1}{\ell}\left( \frac{\epsilon}{2} + \frac{\epsilon^2}{4} \right)
\end{equation}
where we have substituted $\delta M= M\epsilon$. 
For the modes at the lower bound, the angular velocity is modified as.
\begin{equation}
\Omega_{\rm{mod}}=\Omega +\Delta \Omega=\frac{1}{\ell} \left(1-\frac{\epsilon}{2}+\frac{3}{4}\epsilon^2 \right)
\label{omegamodi}
\end{equation}
We proceed by calculating $\Omega_{\rm{mod}} \delta J$ (to second order) for the lower bound in (\ref{range}) to check whether it satisfies the modified form of the Needham's condition given in (\ref{needhammod}).
\begin{eqnarray}
\Omega_{\rm{mod}} \delta J &=& \frac{1}{\ell} \left(1-\frac{\epsilon}{2}+\frac{3}{4}\epsilon^2 \right) \delta M \ell \left(1+ \frac{\epsilon}{2} -\frac{\epsilon^2}{2} \right) \nonumber \\
&=& \delta M
\end{eqnarray}
The lower limit for $\delta J$ for overspinning to occur saturates the modified form of the Needham's condition. In other words, when the angular velocity of the event horizon increases, the lower limit in (\ref{range}) becomes the optimal perturbation. The modes with a higher angular momentum contribution will not be absorbed by the black hole. Therefore the absorption of the challenging modes in the range (\ref{range}) is prevented  and the overspinning problem is fixed. The induced increase in the angular velocity of the horizon has the exact magnitude required to prevent overspinning, which is an interesting fact one could not anticipate at the beginning.
\section{The case of fields}
In \cite{btz} we have also analysed the interaction of BTZ black holes with massless test fields and derived the same result with test bodies, i.e. overspinning is possible for nearly extremal black holes. Though the use of Needham's condition (\ref{needham}) allows us to treat test bodies and test fields almost on equal footage, it is useful to execute an independent analysis for test fields. We start with an incident wave mode of the form:
\begin{equation}
\varphi(r,t,\phi)=e^{-i\omega t}e^{im\phi } f(r)
\end{equation}
where $\omega$ denotes the angular frequency and $m$ denotes the azimuthal wave number as usual. The separation of variables is possible since the BTZ metric admits the Killing vectors $\partial /\partial t$ and $\partial /\partial \phi$. The contribution of such a mode to mass and angular momentum parameters are related by 
\begin{equation}
\delta J=\frac{m}{\omega} \delta M
\label{momega}
\end{equation}
We start with a nearly extremal BTZ black hole which satisfies (\ref{param1}). At the end of the interaction we require that 
\begin{equation}
J+\delta J>(M+\delta M)\ell
\label{field1}
\end{equation}
so that the black hole is overspun into a naked singularity. By imposing (\ref{param1}), (\ref{field1}) implies
\begin{equation}
\delta J-\delta M \ell > M \ell \frac{\epsilon^2}{2}
\label{field2}
\end{equation}
By substituting $\delta J$ from (\ref{momega}) to (\ref{field2}) we derive the upper limit for $\omega$ for overspinning to occur
\begin{equation}
\omega < \frac{m}{\ell (1+\epsilon /2)}
\end{equation}
where we have let $\delta M=M \epsilon$. Note that the upper limit for $\omega$ corresponds to the lower limit for $\delta J$, and vice versa. We derive the upper limit for $\delta J$ or the lower limit for $\omega$ by demanding that the test field is absorbed by the black hole, i.e. the Needham's condition is satisfied. Note that, for a test field satisfying (\ref{momega}), Needham's condition (\ref{needhamneut}) reduces to
\begin{equation}
\omega \geq m\Omega
\label{needhamsuperrad}
\end{equation}
which is the well-known superradiance condition satisfied by bosonic test fields the energy-momentum tensor of which satisfy the null energy condition. The fields with lower frequencies are reflected back to infinity with a larger amplitude, i.e. no net absorption of these fields occur. In a more subtle approach one can say that the absorption probability of these fields is negative. The superradiance condition (or equivalently Needham's condition) implies 
\begin{equation}
\omega \geq \frac{m}{\ell} \left(1-\epsilon+\frac{\epsilon^2}{2} \right)
\end{equation}
where we have substituted the expression for $\Omega$ from (\ref{omegah}).
We have derived the lower and the upper limit for $\omega$ by simultaneously demanding that the test field is absorbed by the black hole and the black hole is overspun, respectively. This determines the range of frequencies that can be used to overspin the black hole:
\begin{equation}
\frac{m}{\ell} \left(1-\epsilon+\frac{\epsilon^2}{2} \right) \leq \omega < \frac{m}{\ell (1+\epsilon /2)}
\label{rangefield}
\end{equation}
Now, we incorporate the induced increase in the angular velocity of the event horizon to test whether overspinning can be prevented. The test fields have to satisfy the modified form of the Needham's condition:
\begin{equation}
\omega \geq m\Omega_{\rm{mod}}
\label{needhamfieldmod}
\end{equation}
where $\Omega_{\rm{mod}}=\Omega +\delta \Omega$ as in the case of test bodies. We want to check whether $\Omega_{\rm{mod}}$ can be sufficiently large to prevent the absorption of the fields with frequencies in the range (\ref{rangefield}). A slight increase in $\Omega$ will prevent the absorption of the modes in the lower limit. We focus on the upper limit for $\omega$ which corresponds to the lower limit for $\delta J$. For these modes:
\begin{equation}
\delta J=\frac{m}{\omega} \delta M=M\ell \left( \epsilon + \frac{\epsilon^2}{2} \right)
\end{equation}
where we have substituted $\delta M=M\epsilon$ and derived the identical result with the case of test bodies. The calculation of the modified value of $\Omega$ gives the same result as (\ref{omegamodi}).
\begin{equation}
\Omega_{\rm{mod}}=\frac{1}{\ell} \left(1-\frac{\epsilon}{2}+\frac{3}{4}\epsilon^2 \right)
\label{omegamodifield}
\end{equation}
We compare the quantity $m \Omega_{\rm{mod}}$ with the upper limit of the range (\ref{rangefield}) to check whether the modified form of the Needham's condition (\ref{needhamfieldmod}) is satisfied. We observe that
\begin{equation}
\frac{m}{\ell (1+\epsilon /2)} < m\Omega_{\rm{mod}}=\frac{1}{\ell} \left(1-\frac{\epsilon}{2}+\frac{3}{4}\epsilon^2 \right)
\end{equation}
To second order the upper limit of the range (\ref{rangefield}) is less than $m \Omega_{\rm{mod}}$, therefore these modes fail to satisfy the modified form of the Needham's condition or equivalently the super-radiance condition.
In other words the modes at the upper limit will be subject to super-radiance as the angular velocity of the event horizon increases. For the frequencies lower than the upper limit, $\delta J$ will be higher and $\Omega_{\rm{mod}}$ will have an even larger value. Therefore we can conclude that the induced increase in the angular velocity precludes the absorption of all the challenging modes and fixes the overspinning problem by modifying the Needham's condition or equivalently the super-radiance condition.
\section{Summary and discussions}
The naked singularities in AdS space-times correspond to states rotating faster than light in the Ads/CFT correspondence. Therefore one would like to prove that their existence is prohibited in the spirit of the cosmic censorship conjecture in classical general relativity.
Previously Rocha-Cardoso evaluated the possibility to overspin a BTZ black hole with test bodies \cite{rc}. Their analysis was restricted to extremal black holes and they derived that extremal BTZ black holes cannot be overspun. They also stated that a process that leads to overspinning of BTZ black holes should not be taken seriously. Following this work, we adapted the same line of research and showed that overspinning is possible if we start with a nearly extremal black hole \cite{btz}. This overspinning is actually not generic. It would be fixed by employing backreaction effects such as gravitational radiation and self-energy, if we were in (3+1) dimensions. However, in (2+1) dimensions, overspinning becomes a drastic problem as the gravitational-wave degrees of freedom are suppressed and the concept of self energy is controversial. 

To bring a solution to this notorious problem we considered the induced increase in the angular velocity of the event horizon formulated by Will \cite{will}. The perturbations satisfying the null energy condition obey Needham's condition (\ref{needham}). The induced increase in the angular velocity modifies Needham's condition and prevents the absorption of the challenging modes. We derived the induced increase in the angular velocity for BTZ space-time, depending on the contribution of the perturbation to the angular momentum parameter ($\delta J$). We determined the range of the possible values of $\delta J$ that could lead to overspinning of BTZ black holes by test bodies. The upper limit for $\delta J$ saturates the Needham's condition. A slight increase in the angular velocity would prevent its absorption. We showed that the induced increase in the angular velocity of the event horizon has the exact value to prevent the absorption of the perturbations at the lower limit of the range. For higher values of ($\delta J$), the induced increase in the angular velocity would be larger. This implies that the challenging modes will not be absorbed by the BTZ black hole as the angular velocity of the event horizon increases. For test fields, Needham's condition is equivalent to the super-radiance condition. We derived the range of frequencies for scalar test fields that could overspin a  nearly extremal BTZ black hole. The upper limit of the frequency corresponds to the lower limit of $\delta J$, as they are inversely proportional. We derived that the upper limit of the challenging frequencies will be subject to super-radiance as the angular velocity of the event horizon increases. Thus, we have brought a solution to the overspinning problem without invoking self-energy or gravitational radiation.

Recently, we have incorporated the explicit form of the absorption probabilities to the scattering problem of test fields \cite{absorp}. From Kerr analogy, we expect the absorption probability for bosonic fields to involve a term $(\omega-m\Omega)^n$ where $n$ is an odd integer. The absorption probability for the super-radiant modes is negative and the probability is zero at the super-radiance limit $\omega=m\Omega$. These modes are entirely reflected back to infinity leaving the space-time parameters invariant. Therefore we can shift the lower limit of the frequency of test fields to higher values, leading to a narrower range. This does not contradict with the result that the induced increase in the angular velocity prevents the absorption of the challenging modes, but reinforces it.

We would like to note that the results derived in this work apply to perturbations satisfying the null energy condition. In that case the Needham's condition is satisfied. There exists a lower limit for $\omega$ which corresponds to an upper limit for $\delta J$ to allow the absorption of the test field. Overspinning due to fermionic fields derived in  \cite{duztas,toth,generic,spinhalf,threehalves}, remain as a separate problem, the solution of which will probably involve the incorporation of quantum effects beyond the semi-classical level.

%

\end{document}